\documentclass[epj]{svjour}

\usepackage{graphicx}
\usepackage{fancyhdr}
\def\makeheadbox{{%  remove EPJC header
 }}
\def\mytitle{My title} 
\def\myauthors{My name}  
\def\mytype{My type of session}
\def\mysession{My session}
\def\mytitle{Impact of slepton generation mixing on the search for sneutrinos} %Put your title here!
\def\myauthors{K. Hidaka}    %Put your name here!
\def\mytype{Contributed Talk}    
\def\mysession{Colliders - SUSY Phenomenology}
\def\beq   {\begin{equation}}
\def\eeq   {\end{equation}}
\def\beqd  {\begin{displaymath}}
\def\eeqd  {\end{displaymath}}
\def\beqaa {\begin{eqnarray}}
\def\eeqaa {\end{eqnarray}}

\def\noi {\noindent}

\def\ti  {\tilde}

\def\sf  {\ti f}

\def\sn  {\ti \nu}
\def\sl  {\ti \ell}

\def\nt  {\tilde\chi^0}
\def\ch  {\tilde\chi^\pm}
\def\chp {\tilde\chi^+}

\def\a   {\alpha}
\def\b   {\beta}
\def\t   {\theta}

\def\sz{\ifmmode{\tilde{\chi}^0} \else{$\tilde{\chi}^0$} \fi}
\def\sw{\ifmmode{\tilde{\chi}} \else{$\tilde{\chi}$} \fi}

\newcommand{\gsim}{\;\raisebox{-0.9ex}
           {$\textstyle\stackrel{\textstyle >}{\sim}$}\;}
\newcommand{\lsim}{\;\raisebox{-0.9ex}{$\textstyle\stackrel{\textstyle<}
           {\sim}$}\;}

\begin{document}
\title{Impact of slepton generation mixing on the search for sneutrinos}
\subtitle{Lepton flavour violation in sneutrino production and decays 
in the general MSSM}
\author{K. Hidaka%\inst{1}
% \thanks is opti onal - remove next line if not needed
%\thanks{\emph{Email:} Insert  Email  of corresponding author here}%
% \and
% Second author\inst{2}% etc
}                     % Do not remove
\institute{Department of Physics, Tokyo Gakugei University, Koganei,
Tokyo 184--8501, Japan
%\and 
}
\date{}
\abstract{
%Insert your abstract here.
We perform a systematic study of sneutrino production and decays in the Minimal 
Supersymmetric Standard Model (MSSM) with slepton generation mixing. 
We study bosonic decays like $\sn \to \sl^-$+ $W^+/H^+$ 
as well as fermionic ones. We show that the effect of slepton generation mixing 
on the sneutrino production and decays 
can be quite large in a significant part of the 
MSSM parameter space despite the very strong experimental limits 
on lepton flavour violating processes. This could have an important impact 
on the search for sneutrinos and the determination of the MSSM parameters 
at future colliders, such as LHC, ILC, CLIC and muon collider.
\PACS{
%      {PACS-key}{discribing text of that key}   \and
%      {PACS-key}{discribing text of that key}
      {12.15.Ji}{Applications of electroweak models to specific processes}   %\and
     } % end of PACS codes
} %end of abstract
\maketitle
%
%------------------------------------------------------------------
\section{Introduction}
%------------------------------------------------------------------
\label{intro}
Systematic studies of decays of sneutrinos, the supersymmetric (SUSY) 
partners of neutrinos, in the Minimal Supersymmetric Standard Model 
(MSSM) have been performed already \cite{Bartl: slepton decay}.
In these studies it is assumed that there is no generation mixing in 
the slepton sector. 
%In these studies it has been assumed that individual lepton flavour is 
%conserved in the slepton sector. 
In this article based on \cite{Bartl: LFVsnu} we study the effect of 
slepton generation mixing on the production and decays of the sneutrinos 
in the MSSM. 
Lepton flavour violating (LFV) productions and decays of SUSY particles 
have been studied for the case of slepton generation mixing \cite{LFV_refs}. 
Some of the studies are rather model dependent. Furthermore, so far no 
systematic study of LFV in sneutrino decays including bosonic decays 
has been performed. The aim of this article is to perform a systematic study 
of sneutrino production and decays including the bosonic decay modes in 
the general MSSM with LFV in slepton sector. 

%------------------------------------------------------------------
\section{The model}
%------------------------------------------------------------------
\label{model}
  First we summarize the MSSM parameters in our analysis. 
The most general charged slepton mass matrix including left-right mixing
as well as flavour mixing in the basis of
$\sl_{0\a}=(\tilde e_L,\tilde\mu_L,\tilde\tau_L,\tilde
e_R,\tilde\mu_R,\tilde\tau_R)$, $\a=1,...,6$,  
is given by \cite{Bartl: LFVsnu}:
%is given by \cite{Bartl: LFV@LHC,Chung}:
%
%\begin{equation}
\[
M^2_{\tilde \ell} = \left(
\begin{array}{cc}
M^2_{LL} &  M^{2\dagger}_{RL} \\
M^2_{RL} &  M^2_{RR} \\
\end{array} \right)~,
%\label{eq:sleptonmass}
\]
%\end{equation}
%
%where the entries are $3 \times 3$ matrices. They are given by
with
\begin{eqnarray*}
%\label{eq:sleptonmassLL}
M^2_{LL,\alpha\beta} &=& 
M^2_{L,\alpha\beta} + m^2_Z\cos(2\b)(-\frac{1}{2}+\sin^2{\t_W})\delta_{\a\b} \\
                     &+&m^2_{\ell_\a}\delta_{\a\b} \, ,\\
%
%\label{eq:sleptonmassRR}
M^2_{RR,\a\b} &=& M^2_{E,\a\b}-m^2_Z\cos(2\b)\sin^2{\t_W}\delta_{\a\b} 
+m^2_{\ell_\a}\delta_{\a\b} \, ,\\
%
%\label{eq:sleptonmassRL}
M^2_{RL,\a\b} &=& v_1 A_{\b\a}-m_{\ell_\a}\mu^*\tan\b\delta_{\a\b} \, .
\end{eqnarray*}
The indices $\a,\b=1,2,3$ characterize the flavours $e,\mu,\tau$, respectively.
$M^2_{L}$ and $M^2_{E}$ are the hermitean soft SUSY breaking mass matrices for
left and right sleptons, respectively. 
$A_{\a\b}$ are the trilinear soft
SUSY breaking couplings of the sleptons and the Higgs boson:
${\mathcal L}_{\rm int}=-A_{\a\b} \sl_{\b R}^\dagger \sl_{\a L} H^0_1 
 + A_{\a\b} \sl_{\b R}^\dagger \sn_{\a L} H^-_1 + \cdots$.
$\mu$ is the higgsino mass parameter.
$v_1$ and $v_2$ are the vacuum expectation values of the Higgs fields
with $v_1=\langle H^0_1\rangle$, $v_2=\langle H^0_2\rangle$, 
and $\tan\b\equiv v_2/v_1$. 
We work in a basis where the Yukawa coupling
matrix $Y_{E,\a\b}$ of the charged leptons is real and flavour
diagonal with $Y_{E,\a\a}=m_{\ell_\a}/v_1=\frac{g}{\sqrt{2}}
\frac{m_{\ell_\a}}{m_W\cos\b}~
(\ell_\a=e,\mu,\tau)$, with $m_{\ell_\a}$ being the physical lepton masses 
and $g$ the SU(2) gauge coupling.
The physical mass eigenstates $\sl_i$, $i=1,...,6$, are given by 
$\sl_i = R^{\sl}_{i\a} \sl_{0\a}$. 
The mixing matrix $R^{\sl}$ and the physical mass eigenvalues
are obtained by an unitary transformation 
$R^{\sl}M^2_{\sl}R^{\sl\dagger}=
{\rm diag}(m^2_{\sl_1},\dots,m^2_{\sl_6})$, 
where $m_{\sl_i} < m_{\sl_j}$ for $i<j$.
Similarly, the mass matrix for the sneutrinos, in the basis
$\sn_{0\a}=(\tilde\nu_{eL},\tilde\nu_{\mu L},\tilde\nu_{\tau L})\equiv
(\tilde\nu_{e},\tilde\nu_{\mu},\tilde\nu_{\tau})$, reads
\begin{eqnarray*}
M^2_{\sn,\a\b} &=&  M^2_{L,\a\b} 
+ \frac{1}{2} m^2_Z\cos(2\b)\delta_{\a\b} 
\quad (\alpha,\beta=1,2,3)~,
%\label{eq:sneutrinomass}
\end{eqnarray*}
where the physical mass eigenstates are given by
$\sn_i = R^{\sn}_{i\a}\sn_{0\a}$, $i=1,2,3$,
($m_{\sn_1} < m_{\sn_2} < m_{\sn_3}$).\\
The properties of the charginos $\ch_i$ ($i=1,2$, $m_{\ch_1}<m_{\ch_2}$) 
and neutralinos $\nt_k$ ($k=1,...,4$, $m_{\nt_1}< ...< m_{\nt_4}$)  
are determined by the parameters $M_2$, $M_1$, $\mu$ and $\tan\b$, 
where $M_2$ and $M_1$ are the SU(2) and U(1) gaugino masses, respectively. 
Assuming gaugino mass unification we take $M_1=(5/3)\tan^2\t_W M_2$. 

\noi
The possible fermionic and bosonic two-body decay modes of sneutrinos are 
\begin{eqnarray*}
  \sn_i  &\longrightarrow & \nu \nt_j,~ \ell^-_\a \chp_k~, \\
   \sn_i  &\longrightarrow & \sl^-_j W^+,~ \sl^-_j H^+~.    %\label{eq:Bmode}
\end{eqnarray*}

%------------------------------------------------------------------
\section{Constraints}
%------------------------------------------------------------------
\label{constraints}
In our analysis, we impose the following conditions on the MSSM parameter space 
in order to respect experimental and theoretical constraints which are described in detail in \cite{Bartl: LFVsnu}:

\renewcommand{\labelenumi}{(\roman{enumi})} 
% set counter to small roman numbers
\begin{enumerate}
  \item The vacuum stability conditions \cite{Dimopoulos}, such as \\
        $|A_{\a\b}|^2 < Y_{E,\gamma\gamma}^2(M^2_{L,\a\a} + M^2_{E,\b\b} + m^2_1)$,
        ($\a\neq\b$; $\gamma=Max(\a,\b)$; $\a,\b=1,2,3=e,\mu,\tau$). 
  \item Experimental limits on the LFV lepton decays: \\
        $B(\mu^- \to e^- \gamma) < 1.2 \times 10^{-11}$ (90\% CL) 
                                   \cite{Brooks},
        $B(\tau^- \to \mu^- \gamma) < 4.5 \times 10^{-8}$ (90\% CL) 
                                      \cite{Belle: ICHEP2006},
        $B(\tau^- \to e^- \gamma) < 1.1 \times 10^{-7}$ (90\% CL) 
                                    \cite{Babar: 2006},
        $B(\mu^- \to e^- e^+ e^-) < 1.0 \times 10^{-12}$ (90\% CL) 
                                    \cite{SINDRUM},
        $B(\tau^-\to \mu^-\mu^+\mu^-) < 3.2 \times 10^{-8}$ (90\% CL) 
                                        \cite{BELLE: tau to 3l},
        $B(\tau^-\to e^-e^+e^-) < 3.6 \times 10^{-8}$ (90\% CL) 
                                        \cite{BELLE: tau to 3l}.
  \item Experimental limits on SUSY contributions to \\
        anomalous magnetic moments of leptons \cite{PDG,Jegerlehner}
\footnote{
      For the limit on SUSY contributions to 
      anomalous magnetic moment
      of muon $\Delta a_\mu^{SUSY}$, we allow for 
      an error at 95\% CL for the difference 
      between the experimental measurement and the SM prediction 
      \cite{Jegerlehner}.}, e.g.
        $|\Delta a_\mu^{SUSY} - 287 \times 10^{-11}|<178 \times 
        10^{-11}$ (95\% CL).
  \item The LEP limits on SUSY particle masses. 
  \item The limit on $m_{H^+}$ and $\tan\b$ from the experimental data on 
        $B(B_u^- \to \tau^- {\bar\nu}_\tau)$ \cite{BELLE: Btaunu}.

\end{enumerate}
It has been shown that in general the limit on the
$\mu^-- e^-$ conversion rate is respected if the
limit on $\mu\to e~ \gamma$ is fulfilled \cite{Hisano:1995cp}.\\
Condition (i) strongly constrains the trilinear couplings
$A_{\alpha\beta}$, especially for small $\tan\beta$ where
the lepton Yukawa couplings $Y_{E,\a\a}$ are small.  
(ii) strongly constrains the lepton flavour mixing parameters; e.g. 
in case of $\tilde\mu-\tilde\tau$ mixing the limit on $B(\tau^- \to \mu^- \gamma)$ 
strongly constrains the $\tilde\mu-\tilde\tau$ mixing parameters $M^2_{L,23}, 
M^2_{E,23}$, $A_{23}$ and $A_{32}$. 
%The limit on SUSY contributions to anomalous magnetic moment of muon 
%in (iii) is also important, e.g. it disfavours negative $\mu$ especially 
%for large $\tan\b$. 
The limit on $\Delta a_\mu^{SUSY}$ 
in (iii) is also important, e.g. it disfavours negative $\mu$ especially 
for large $\tan\b$. 

%------------------------------------------------------------------
\section{Numerical results}
%------------------------------------------------------------------
\label{numerics}
We take $\tan\b, m_{H^+}, M_2, \mu, M^2_{L,\a\b}, M^2_{E,\a\b}$, and $A_{\a\b}$ 
as the basic MSSM parameters at the weak scale. We assume them to be real. 
The LFV parameters are $M^2_{L,\a\b}$, $M^2_{E,\a\b}$, and $A_{\a\b}$ with $\a \neq \b$.
We take the following $\ti\mu-\ti\tau$ mixing scenario as a reference scenario with
LFV within reach of LHC and ILC: \\
$\tan\b=20$, $m_{H^+}=150GeV$, $M_2=650GeV$, $\mu=150GeV$, 
$M^2_{L,11}=(430GeV)^2$, $M^2_{L,22}=(410GeV)^2$, $M^2_{L,33}=(400GeV)^2$, 
$M^2_{L,12}=M^2_{L,13}=(1GeV)^2$, $M^2_{L,23}=(61.2GeV)^2$, 
$M^2_{E,11}=(230GeV)^2$, $M^2_{E,22}=(210GeV)^2$, $M^2_{E,33}=(200GeV)^2$, 
$M^2_{E,12}=M^2_{E,13}=(1GeV)^2$, $M^2_{E,23}=(22.4GeV)^2$,
$A_{23}=25GeV, A_{33}=150GeV$, and all the other $A_{\a\b}=0$.\\
In this scenario satisfying all the conditions (i)-(v) above we have:\\
$
m_{\sn_1}=393GeV, m_{\sn_2}=407GeV, m_{\sn_3}=425GeV, \\
\vspace{-0.3cm}\\
\sn_1=-0.36\sn_{\mu}+0.93\sn_{\tau} \sim \sn_{\tau},\\
\sn_2=0.93\sn_{\mu}+0.36\sn_{\tau} \sim \sn_{\mu},\\
\sn_3 \simeq \sn_{e}, \\
\vspace{-0.3cm}\\
m_{\sl_1}=204GeV, m_{\sl_2}=215GeV, m_{\sl_3}=234GeV, \\
\vspace{-0.3cm}\\
\sl_1= -0.0029\ti{\mu}_L+0.033\ti{\tau}_L
-0.12\ti{\mu}_R+0.99\ti{\tau}_R \sim \ti{\tau}_R,\\
\sl_2=0.0022\ti{\mu}_L+0.0040\ti{\tau}_L
+0.99\ti{\mu}_R+0.12\ti{\tau}_R \sim \ti{\mu}_R,\\
\sl_3 \simeq \ti{e}_R,\\
\vspace{-0.3cm}\\
B(\sn_1 \to \mu^- + \chp_1)=0.014, B(\sn_1 \to \tau^- + \chp_1)=0.36, \\
B(\sn_1 \to \sl^-_1 + H^+)=0.48, \\
\vspace{-0.3cm}\\
B(\sn_2 \to \mu^- + \chp_1)=0.20, B(\sn_2 \to \tau^- + \chp_1)=0.12, \\
B(\sn_2 \to \sl^-_1 + H^+)=0.38.
$
\\
\vspace{-0.3cm}\\
As $\sn_2 \sim \sn_\mu$ and $\sl^-_1 \sim \ti\tau^-_R$, the decays 
$\sn_2 \to \tau^- \chp_1$ and $\sn_2 \to \sl^-_1 H^+$ 
are essentially LFV decays. Note that the branching ratios of these LFV decays 
are sizable in this scenario. The reason is as follows: 
The lighter neutralinos $\nt_{1,2}$ and the lighter chargino $\ch_1$ 
are dominantly higgsinos as $M_{1,2} \gg |\mu|$ in this scenario. 
Hence the fermionic decays into 
$\nt_{1,2}$ and $\chp_1$ are suppressed by the small lepton Yukawa 
couplings except for the decay into $\tau^- \chp_1$ which does not 
receive such a suppression because of the sizable $\tau$ Yukawa coupling 
$Y_{E,33}$ for large $\tan\b$. This leads to an enhancement of 
the bosonic decays into the Higgs boson $H^+$. 
Moreover the decay $\sn_2(\sim \sn_\mu) \to \sl^-_1(\sim \ti\tau^-_R)+ H^+$ 
is enhanced by the sizable trilinear $\sn_\mu-\ti\tau^+_R-H^-_1$ coupling 
$A_{23}$ (with $H^-_1=H^-\sin\b$). 
Because of the sizable $\tilde\nu_\mu-\tilde\nu_\tau$ mixing term $M^2_{L,23}$
the $\sn_2$ has a significant $\sn_\tau$ component, which 
results in a further enhancement of this decay due to the large trilinear 
$\sn_\tau-\ti\tau^+_R-H^-_1$ coupling $A_{33}$ ($=150$~GeV). \\
The decays of $\sn_1$ and $\sn_2$ into $\sl^-_{1,2} W^+$ are suppressed since 
$\sl^-_1 \sim \ti\tau^-_R$ and $\sl^-_2 \sim \ti\mu^-_R$.

%Note that in experimental searches for LFV in sneutrino decays it is important 
%to have at least two different lepton-flavour modes with sizable branching ratios 
%in decay of a sneutrino; e.g. {\it both} sizable $B(\sn_2 \to \mu^- \chp_1)$ 
%{\it and} sizable $B(\sn_2 \to \tau^- \chp_1)$. 

%------------------------------------------------------------------
\subsection{$\tilde\nu$ decay branching ratios }
%------------------------------------------------------------------
\label{BRs}
We study the basic MSSM parameter dependences of the LFV 
sneutrino decay branching ratios for the reference scenario specified above.
In Fig.1 we show contours of the LFV $\sn_2$ decay branching ratios 
in the $\mu-M_2$ plane. All basic parameters other than 
$\mu$ and $M_2$ are fixed as in the reference scenario specified above. 
We see that the LFV decay branching ratios 
$B(\sn_2 \to \tau^-  \chp_1)$ and 
$B(\sn_2 \to \sl^-_1  H^+)$ can be sizable in a significant part of 
the $\mu-M_2$ plane. 
The main reason for the increase of $B(\sn_2\to  \sl^-_1 H^+)$
in the region $M_2\gg \mu$ is that the partial widths for the decays into
$\mu^-\tilde\chi^+_1$ and $\nu\tilde\chi^0_{1,2}$ decrease for increasing 
$|M_2/\mu|$ as the lighter chargino/neutralino states become more and more
higgsino like.
The $\tau^-\tilde\chi^+_1$ decay mode has a different behaviour
due to the sizable $\tau$ Yukawa coupling for large $\tan\beta$.
We remark that the limit on $\Delta a_\mu^{SUSY}$ excludes 
the region with $B(\sn_2 \to \sl^-_1  H^+)\gsim 0.5$. \\
%For the LFV $\tilde\nu_1$ decay we obtain $B(\sn_1 \to
%\mu^-\tilde\chi^+_1)=(0.015,0.03,0.05)$for $(\mu,M_2)$~(GeV) 
%$= (200,600),(400,360),(600,280)$, respectively. \\
\begin{figure}
% For example, with the option graphicx use
\includegraphics[width=0.4\textwidth,height=0.25\textwidth,angle=0]{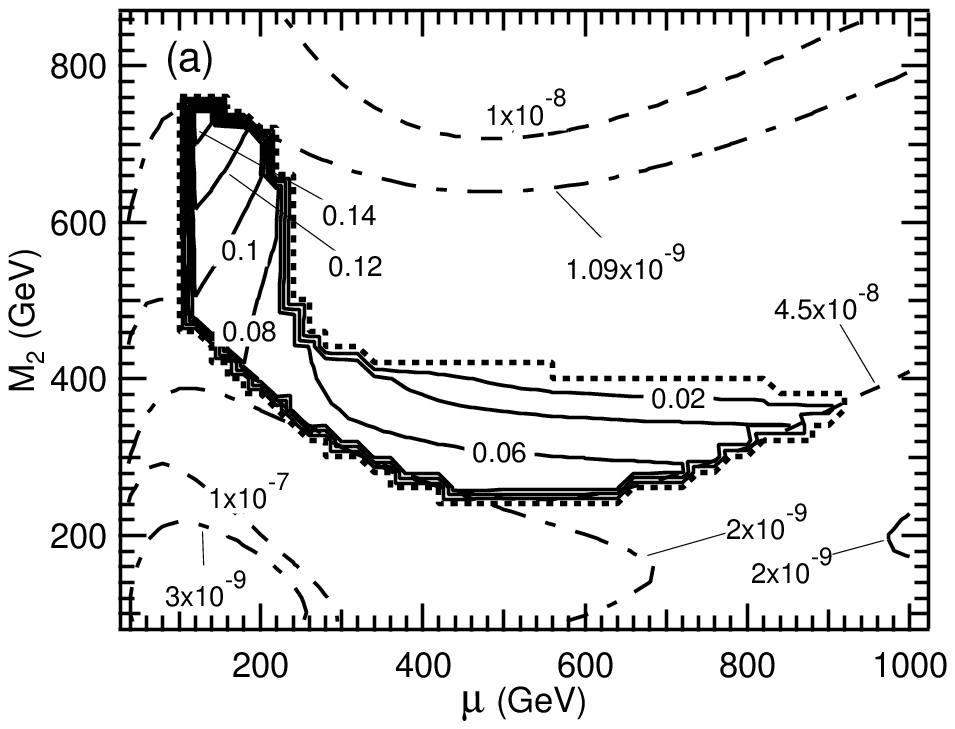}
\includegraphics[width=0.4\textwidth,height=0.25\textwidth,angle=0]{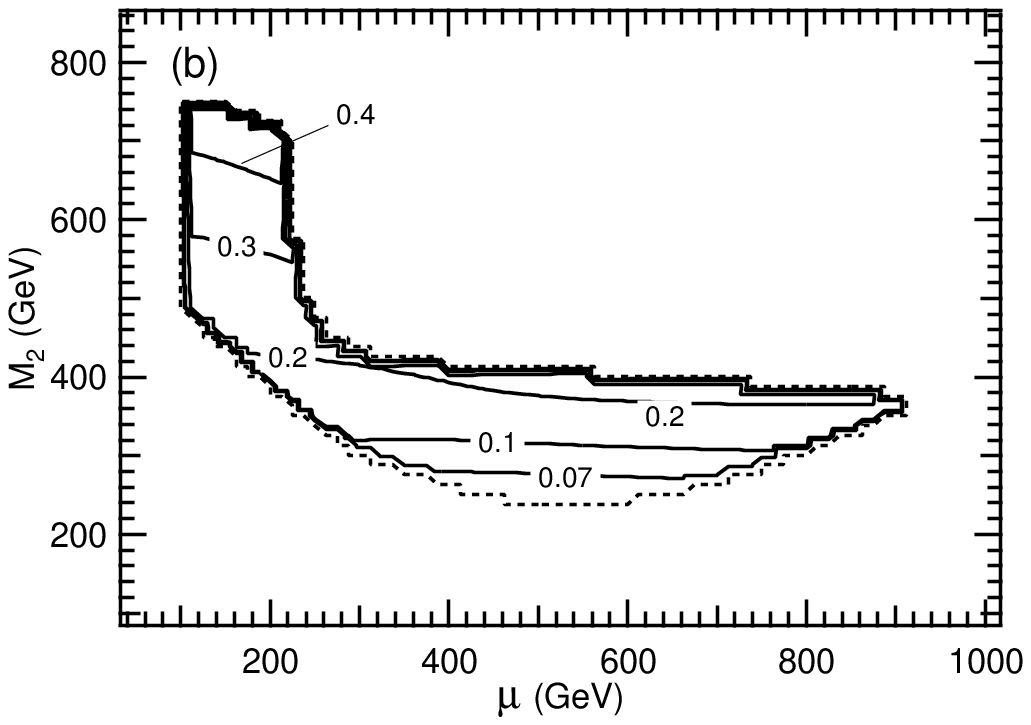}
%\vspace{-0.2cm}
\caption{
Contours of (a) $B(\sn_2 \to \tau^-  \chp_1)$ 
and (b) $B(\sn_2 \to \sl^-_1  H^+)$ in the $\mu-M_2$ 
plane for our $\ti\mu-\ti\tau$ mixing scenario.  
The region with no solid contour-lines is excluded by the conditions (i) to 
(v) given in the text; negative $\mu$ region is excluded by the limit on 
$\Delta a_\mu^{SUSY}$ in (iii). The dashed and dash-dotted lines in (a) 
show contours of $B(\tau^- \to \mu^-  \gamma)$ and $\Delta a_\mu^{SUSY}$, 
respectively. Note that (iii) requires $1.09 \times 10^{-9} < 
\Delta a_\mu^{SUSY} < 4.65 \times 10^{-9}$.
}
\label{fig1}       % Give a unique label
\end{figure}
%
%
%\noi
In the following we use the quantities $R_{L23}\equiv
M^2_{L,23}/$ $((M^2_{L,11}+M^2_{L,22}+M^2_{L,33})/3)$ and $R_{A23} \equiv 
A_{23}/((|A_{11}|+|A_{22}|+|A_{33}|)/3)$ as a measure of LFV.
In Fig.2 we present the $R_{L23}$ dependence of $\sn_2$ decay branching 
ratios, where all basic parameters other than $M^2_{L,23}$ are 
fixed as in the reference scenario specified above. 
We see that the LFV decay branching ratios $B(\sn_2 \to \tau^-  \chp_1)$ 
and $B(\sn_2 \to \sl^-_1  H^+)$ can be large and very sensitive to $R_{L23}$. 
Note that $\sl^-_1 \sim \ti\tau^-_R$ and that the $\sn_\tau$ component in 
$\sn_2(\sim \sn_\mu)$ increases with the increase of the $\ti\nu_\mu - \ti\nu_\tau$ 
mixing parameter $M^2_{L,23}$, which explains the behaviour of the branching ratios.
Similarly we have found that $B(\sn_2 \to \sl^-_1  H^+)$ can be very sensitive 
to $R_{A23}$; this decay can be enhanced also by a sizable $A_{23}$ 
as explained above.
\begin{figure}
% For example, with the option graphicx use
\includegraphics[width=0.4\textwidth,height=0.25\textwidth,angle=0]{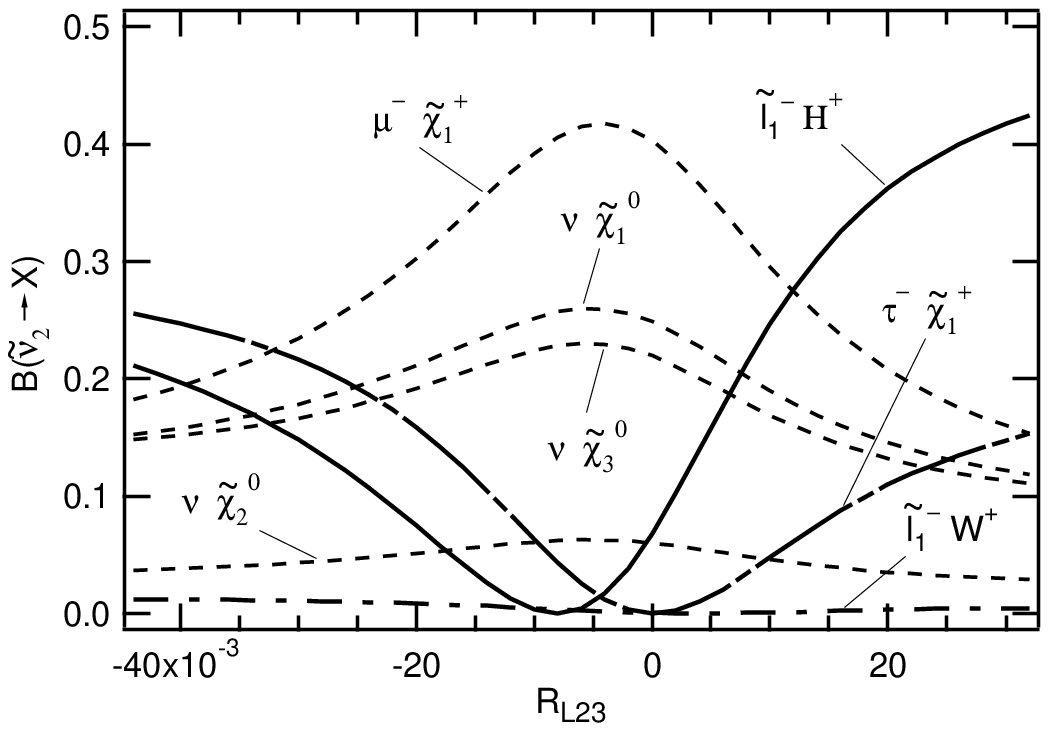}
\caption{
$R_{L23}$ dependence of $\sn_2$ decay branching ratios for our 
$\ti\mu-\ti\tau$ mixing scenario.  The shown range of 
$R_{L23}$ is the whole range allowed by the conditions 
(i) to (v) given in the text.
}
\label{fig2}       % Give a unique label
\end{figure}
To exemplify this behaviour further, in Fig.3 we show the contours
of these decay branching ratios in the $R_{L23}-R_{A23}$ plane, where
all basic parameters other than $M^2_{L,23}$ and $A_{23}$ are fixed
as in the reference scenario specified above.
As can be seen, these LFV decay branching ratios can be
large in a sizable region of the $R_{L23}-R_{A23}$ plane
and their dependences on $R_{L23}$ and $R_{A23}$ are quite remarkable and 
very different from each other.
Hence, a simultaneous measurement of these two branching 
ratios could play an important role in determination of the LFV parameters 
$M^2_{L,23}$ and $A_{23}$. \\
\begin{figure}
% For example, with the option graphicx use
\includegraphics[width=0.40\textwidth,height=0.25\textwidth,angle=0]{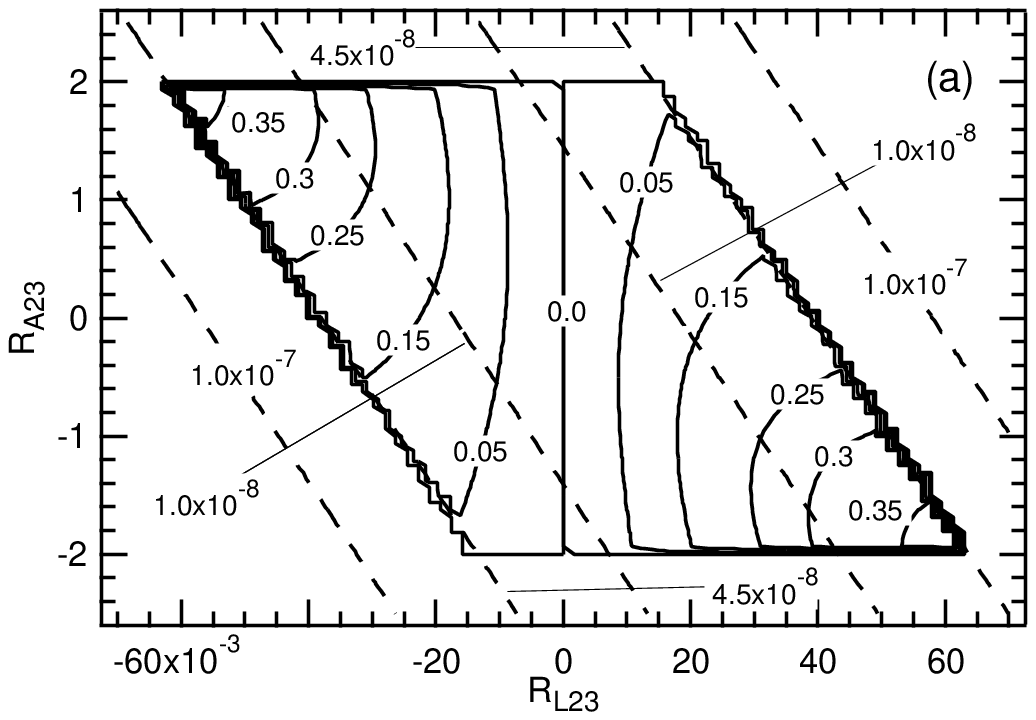}
\includegraphics[width=0.40\textwidth,height=0.25\textwidth,angle=0]{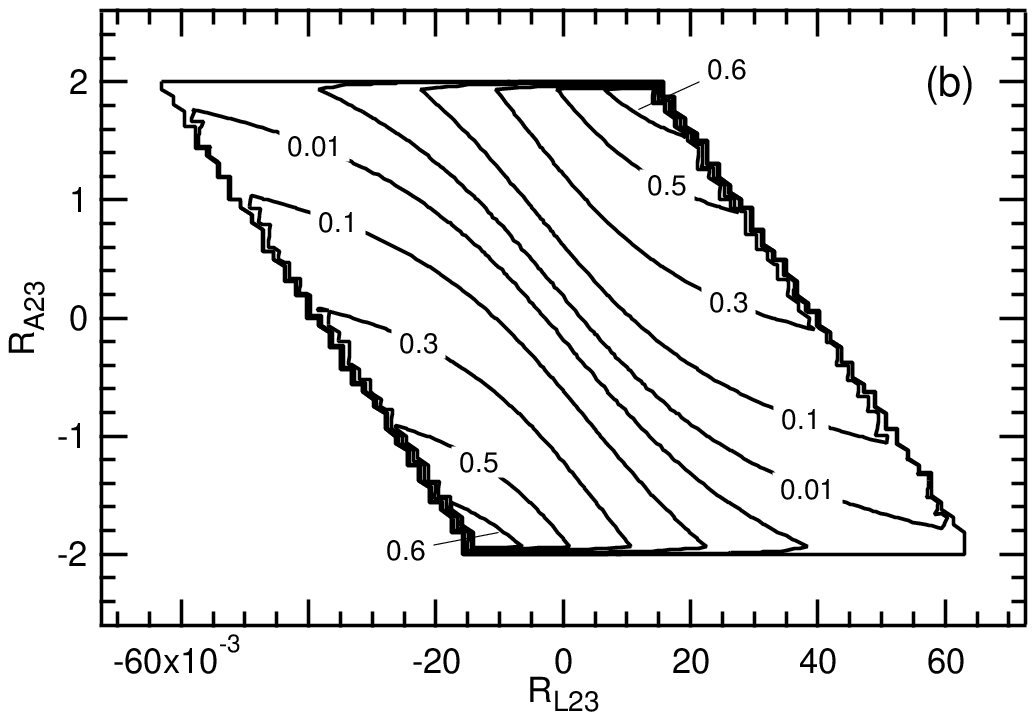}
\caption{
Contours of the LFV decay branching ratios (a) $B(\sn_2 \to \tau^-  \chp_1)$ 
and (b) $B(\sn_2 \to \sl^-_1  H^+)$ 
in the $R_{L23}-R_{A23}$ plane for our $\ti\mu-\ti\tau$ mixing scenario.
The region with no solid contours is excluded by the conditions (i) to (v) 
given in the text. The dashed lines in (a) show contours of 
$B(\tau^- \to \mu^- \gamma)$. 
}
\label{fig3}       % Give a unique label
\end{figure}
In Fig.4 we show a scatter plot of the LFV decay branching ratios 
$B(\sn_2 \to \sl^-_1  H^+)$ versus $B(\tau^- \to \mu^- \gamma)$ for our 
$\ti\mu-\ti\tau$ mixing scenario with the parameters 
$M_2, \mu, R_{L23}, R_{E23}, R_{A23}$ and $R_{A32}$ varied in the ranges 
$0<M_2<1000$~GeV, $|\mu|<1000$~GeV, $|R_{L23}|<0.1, |R_{E23}|<0.2, |R_{A23}|<2.5$ 
and $|R_{A32}|<2.5$, satisfying the conditions (i) to (v) given above. 
All parameters other than $ M_2, \mu, M^2_{L,23}, M^2_{E,23}, 
A_{23} $ and $A_{32}$ are fixed as in the reference scenario specified above.
As can be seen in Fig.4, the LFV branching ratio 
$B(\sn_2 \to \sl^-_1  H^+)$ could go up to 30\%
even if the present bound on $B(\tau^- \to \mu^- \gamma)$ improves
by one order of magnitude.
\begin{figure}
% For example, with the option graphicx use
\includegraphics[width=0.40\textwidth,height=0.25\textwidth,angle=0]{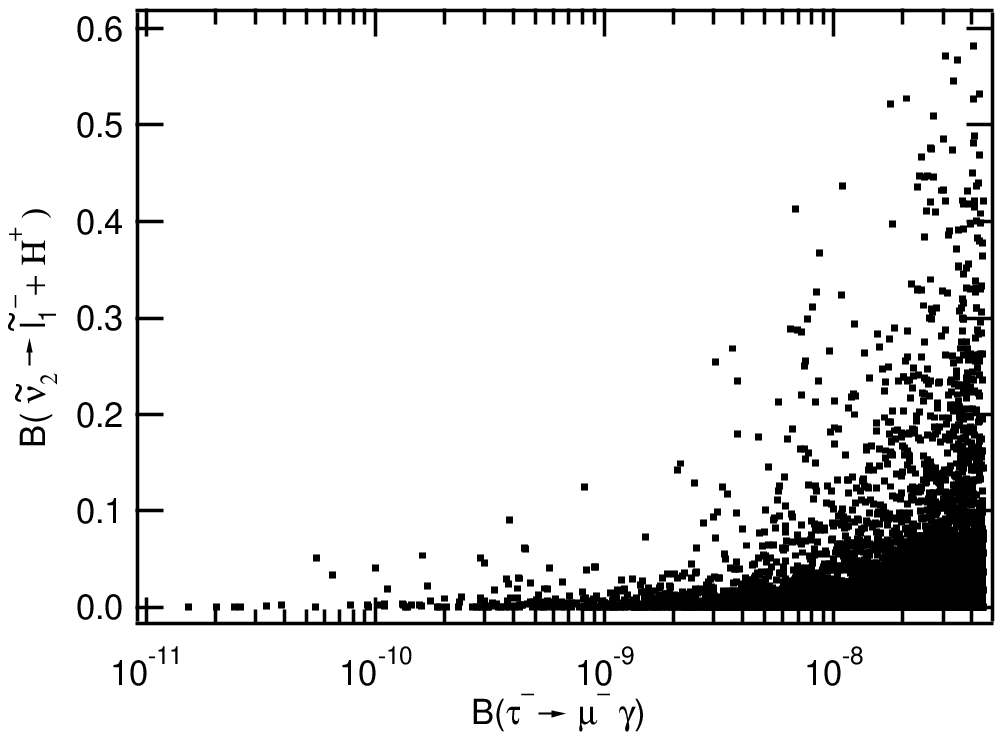}
\caption{
Scatter plot of the LFV decay branching ratios 
$B(\sn_2 \to \sl^-_1  H^+)$ versus $B(\tau^- \to \mu^- \gamma)$ for our 
$\ti\mu-\ti\tau$ mixing scenario. 
}
\label{fig4}       % Give a unique label
\end{figure}
For the other LFV decay branching ratios of $\sn_{1,2}$ versus 
$B(\tau^- \to \mu^-  \gamma)$ we have obtained scatter plots similar to that 
for $B(\sn_2 \to \sl^-_1  H^+)$ versus $B(\tau^- \to \mu^-  \gamma)$, with 
the upper limits of the $\sn_{1,2}$ decay branching ratios 
$B(\sn_1 \to \mu^-  \chp_1) \lsim 0.12$, $B(\sn_1 \to \sl^-_2  H^+) \lsim 0.40$, 
$B(\sn_1 \to \sl^-_2  W^+) \lsim 0.05$, $B(\sn_2 \to \tau^-  \chp_1) \lsim 0.35$, 
and $B(\sn_2 \to \sl^-_1  W^+) \lsim 0.22$.
Note that $B(\sn_1 \to \sl^-_2  H^+)$ can be very large 
due to sizable $M^2_{E,23}$, $A_{32}$ and large $A_{33}$.

We have also studied sneutrino decay branching ratios in the case of 
$\ti e-\tilde\tau$ mixing, where we have obtained similar results to those 
in the case of $\tilde\mu-\tilde\tau$ mixing. 
This is due to the fact that $Y_{E,11} \sim Y_{E,22} \, (\sim 0)$, 
that the experimental limits on $B(\tau^- \to e^- \gamma)$ and 
$B(\tau^- \to \mu^- \gamma)$ are comparable, and that the theoretical limits 
of the condition (i) on the LFV parameters $A_{13}$ and $A_{31}$ are also
similar to those on $A_{23}$ and $A_{32}$.

%------------------------------------------------------------------
\subsection{LFV contributions to collider signatures}
%------------------------------------------------------------------
\label{signatures}
It is to be noted that in $\ti e-\ti\tau$ mixing scenario the 
t-channel chargino exchanges contribute significantly to the cross 
sections $\sigma(e^+e^-\to\sn_i\bar{\sn}_j)\equiv \sigma_{ij}$ for 
$i,j=$1,3, enhancing the cross sections (including the LFV production 
cross section $\sigma_{13}$) strongly, where $\sn_1 \sim \sn_\tau$ 
and $\sn_3 \sim \sn_e$ \cite{Bartl: LFVsnu}.
We have studied the LFV contributions to signatures of sneutrino
production and decay at the ILC \cite{Bartl: LFVsnu}.
We have shown that the LFV processes (including the LFV $\sn_i$ 
productions and the LFV bosonic $\sn_i$ decays also) can contribute 
significantly to signal event rates. 
For example, in the $\tilde e-\tilde\tau$ mixing scenario described in 
\cite{Bartl: LFVsnu}, assuming ILC with $\sqrt{s}=1$~TeV and a 
longidutinal polarization of -90\% and 60\% for the electron and 
positron beam, respectively, 
%the dominant LFV contributions to the rate of the signal 
the dominant LFV contributions (steming from $e^+e^-$ $\to$
$\sn_i\bar{\sn}_j$ $\to$ $e^\pm \tau^\mp \tilde\chi^+_1 \tilde\chi^-_1$)
to the rate of the signal event $e^\pm \tau^\mp+4{\rm jets}+/\!\!\!\!E$ 
is calculated to be $\sigma^{LFV}=6.6$fb, where $/\!\!\!\!E$ is the 
missing energy.
Lepton flavour conserving (LFC) processes in $\sn$ production and decay 
can also contribute to the rate of the signal event above.
%$e^\pm \tau^\mp+4{\rm jets}+/\!\!\!\!E$.
The dominant LFC contributions to the signal rate is calculated to be 
$\sigma^{LFC}=0.033$fb which is two orders of magnitude smaller 
than $\sigma^{LFV}$. 
This strongly suggests that one should take into account the 
possibility of the significant contributions of both the LFV 
fermionic and bosonic decays in the sneutrino search and 
should also include the LFV parameters in 
the determination of the basic SUSY parameters at colliders.
It is clear that detailed Monte Carlo studies taking into account
background and detector simulations are necessary.
However, this is beyond the scope of the present article.

%-------------------------------------------------------------------
\section{Summary}
%-------------------------------------------------------------------
\label{summary}
We have performed a systematic study of sneutrino production and 
decays including both fermionic and bosonic decays in the general MSSM 
with slepton generation mixings. We have shown that LFV sneutrino production 
cross sections and LFV sneutrino decay branching ratios can be quite large 
due to slepton generation mixing in a significant part of the MSSM 
parameter space despite the very strong experimental limits on LFV processes.
This could have an important impact on the search for sneutrinos and the 
MSSM parameter determination at future colliders, such as LHC, ILC, 
CLIC and muon collider.

%------------------------------------------------------------------------
\section*{Acknowledgements}
%------------------------------------------------------------------------
I sincerely thank the other authors of \cite{Bartl: LFVsnu}: A.~Bartl, 
K.~Hohenwarter-Sodek, T.~Kernreiter, W.~Majerotto and W.~Porod.

\end{document}